\documentclass[sigconf]{acmart}

\usepackage{amsmath}
\usepackage{booktabs} % For formal tables
\usepackage{caption}
\usepackage{color}
\usepackage{etoolbox}
\usepackage{graphicx}
\usepackage{multirow}
\usepackage{xspace}
\usepackage{url}

\setcopyright{none}
\begin{document}
\title{AndroVault: Constructing Knowledge Graph from Millions of Android Apps for Automated Analysis}

\author{\large {Guozhu Meng {\normalsize *}, Yinxing Xue {\normalsize \dag}, Jing Kai Siow {\normalsize *}, Ting Su {\normalsize *}, Annamalai Narayanan{\normalsize *},  Yang Liu {\normalsize *}}}

\affiliation{\institution{$^{\normalsize *}$Nanyang Technological University, Singapore\quad$^{\normalsize \dag}$Microsoft, China}}

\settopmatter{printacmref=false} % Removes citation information below abstract
\renewcommand\footnotetextcopyrightpermission[1]{} % removes footnote with conference information in first column
\pagestyle{plain} % removes running headers

%%%%%%%%%%%%%%%%%%%%%%%%%%%%%%%%%%%%%%%%%%%%%%%%%%%%%%%%%%%%%%%%%%%%%%%%%%%
%                   Configuration for this paper                          %
%For example, you can create a command for the name of the framework, some%
%data which is used in many places in the paper                           %
%%%%%%%%%%%%%%%%%%%%%%%%%%%%%%%%%%%%%%%%%%%%%%%%%%%%%%%%%%%%%%%%%%%%%%%%%%%

\newcommand{\tool}{\textsc{AndroVault}\xspace}
\newcommand{\numOfMalware}{1,004,550\xspace}
\newcommand{\numOfDistinct}{4,039,468\xspace}
\newcommand{\numOfAuthor}{232,536\xspace}
\newcommand{\numOfFamily}{1,400\xspace}
\newcommand{\numOfAVClass}{200,373\xspace}
\newcommand{\numOfMarkets}{28\xspace}
\newcommand{\numOftrdMarkets}{27\xspace}
\newcommand{\numOfSources}{33\xspace}

\newcommand{\numOfApps}{5,005,669\xspace}
\newcommand{\numOfAttributes}{36\xspace}
\newcommand{\numOfFacts}{3\xspace}
\newcommand{\numOfRelations}{13\xspace}
\newcommand{\numOfMetaData}{16\xspace}
\newcommand{\numOfJava}{8K\xspace}
\newcommand{\numOfPython}{15K\xspace}%14,513
\newcommand{\numOfAppDesc}{1,180,141\xspace}
\newcommand{\numOfOtherApps}{2,759,053\xspace}
\newcommand{\numOfPackages}{27,560\xspace}
\newcommand{\numOfAppInPackages}{71,888\xspace}
\newcommand{\numOfVersions}{2.61\xspace}
\newcommand{\numOfPiggybackedApps}{21,037\xspace}
\newcommand{\numOfCorruptedApps}{10,348\xspace} %confirm later
\newcommand{\numOfMethods}{7,312\xspace} %confirm later

\begin{abstract}
Data driven research on Android has gained a great momentum these years. The abundance of data facilitates knowledge learning, however, also increases the difficulty of data preprocessing. Therefore, it is non-trivial to prepare a demanding and accurate set of data for research. In this study, we put forward \tool, a framework for the Android research composing of data collection, knowledge representation and knowledge extraction. It has started with a long-running web crawler for data collection (both apps and description) since 2013, which guarantees the timeliness of data; With static analysis and dynamic analysis of the collected data, we compute a variety of attributes to characterize Android apps. After that, we employ a knowledge graph to connect all these apps by computing their correlation in terms of attributes; Last, we leverage multiple technologies such as logical inference, machine learning, and correlation analysis to extract facts % (more accurate and demanding, either high level or not, data)
that are beneficial for specific research problems. Based on the produced data of high quality, we have successfully conducted many research works including malware detection, malware spread, and Android testing. We would like to release our data to the research community in an authenticated manner, and encourage them to conduct productive research.

\end{abstract}

% \fi
%\IEEEpeerreviewmaketitle

\maketitle

\section{Introduction}

Android becomes the biggest mobile platform since 2010. To date, Android further consolidates its domination in the mobile world with reaching the largest market share of 86.8\%~\cite{idc2016}. An enormous number of Android applications (referred as to ``apps'' hereinafter) provide a great convenience to Android users to conduct network life like financial management, online shopping, online eduction, entertainment, etc. These apps help to construct a huge Android ecosystem in which everyone can complete their online activities.

In the past years, Android has gained a remarkable attraction from researchers of both academia and industry. There are five popular research directions on Android: malware detection~\cite{flowdroid2014,amandroid2014,drebin2014,copperdroid2015,iccta2015,smart2016,anna_ijcnn,mamadroid2017,jitana2017}, app testing~\cite{swifthand2013,trimdroid2016,sapienz2016,stoat2017}, clone detection~\cite{dnadroid2012,3dcfg2014,wukong2015}, advertisement analysis~\cite{phalib2016,backes2016,rastogi2016,libd2017}, and performance analysis~\cite{perfchecker2014,clapp2015,dune2016,defdroid2016}. All of these studies rely on a massive number of apps for analysis, and to some extent, the qualify of data. % that is, ``Big Data'' has gained the popularity in the Android domain.
Android apps of a larger number may be more representative and informative, and thereby more approximate to the matters in the real world. However, one fine-grained data set is desired to conduct a comprehensive study on specific research problems such as malware detection.

%\textcolor{red}{Encountered problems in research}
Along with the drastic increase of data and computing power, the research on Android has migrated from methodology driven to data driven that depends on the quality and scale of data to a large extent~\cite{challenge2017}. There are emerging a variety of data driven research works which learn knowledge from a large scale of data, and then complete specific research tasks based on the knowledge~\cite{massvet2015,backes2016,li2017understanding}. Apparently, the quality of data is important to the data driven research since it determines the quality of extracted knowledge. According to~\cite{roebuck2011data,askham2013six}, the quality of data is mainly referred to as the completeness, consistency, timeliness and accuracy of data for a specific purpose. However, many research works are still using outdated data such as \textsc{MalGenome}~\cite{malgenome2012} to learn malware features, or using insufficient data that may produce inaccurate results.
In order to provide ample data of high quality for Android research, we have been dedicated to a continuous effort to collect Android apps and their descriptive information from a wide variety of sources.

To fulfil a specific research task on the data of an enormous volume, researchers have to perform data preprocessing, for instance, data cleaning, clustering and filtering. However, the volume of data can raise the difficulty of data preprocessing, and consequently knowledge learning. \textsc{AndroZoo}~\cite{androzoo2016} provides the largest number of Android apps for the community. However, it is extremely arduous and frustrating to select the data of the most interest for research. To tackle this problem, we compute \numOfAttributes attributes (see Section~\ref{sec:method:attribute} for detail) to characterize apps, and construct a comprehensive knowledge graph by establishing connections between all Android apps in terms of these attributes as detailed in Section~\ref{sec:method:graph}. Given a variety of relationships between entities in the knowledge graph, it becomes much easier yet faster to extract data of common interest from millions of apps.

To further expedite research works, we incorporate some technologies in \emph{Knowledge Vault}~\cite{knowledgevault2014} to obtain more mature and abstract data, a.k.a, \emph{facts} from the graph. Analogy to Knowledge Vault fusing structured facts in \emph{Knowledge Graph}~\cite{knowledgegraph2013} with learned facts by supervisor machine learning, \tool has extract \numOfFacts types of facts that could be directly used for research. In this study, facts are more abstract and composite properties that are put on one or more apps. For example, we define a fact that the apps with same package name and version code but different signing certificates are likely piggybacked. Hence, we can obtain a relatively small but more accurate data set for the app piggybacking study (More details are referred to in Section~\ref{sec:method:fact}). %and produce a list of apps of the same developer ranging in a period of time to facilitate an evolutionary study.
As a consequence, it can greatly reduce the effort to select and process apps from the massive number of apps.

To date, \tool has powered up several research studies including malware detection~\cite{smart2016,anna_ijcnn,onlinelearning2017}, malware generation~\cite{mystique2016,mystiques2017}, Android app testing~\cite{stoat2017}, malware spread study, vulnerability detection, and auto GUI code generation. Fortunately, it is a growing and extendable data warehouse: \tool is continuously collecting more apps and meta data from the Web; its knowledge graph can be extended by computing new attributes, and building new relationships between entities; more sophisticated computations can be conducted to extract required facts.

%\textcolor{red}{How we address this}
%\emph{Knowledge　Vault} \cite{knowledgevault2014} is a web-scale approach to construct a probabilistic knowledge on the basis of \emph{Knowledge Graph}~\cite{knowledgegraph2013}. It extracts information triples from Web, and employs supervised machine learning methods to fuse with structured facts in Knowledge Graph. The probabilities in the base denote the likelihood of fact correctness.

\noindent\textbf{Contributions.} We make the following contributions.

\begin{itemize}
  \item \emph{Vigorous and continuous endeavors in collecting and processing Android apps.} The \textsc{AndroVault} work dates back to 2013, and ranges in the period of Android's rapid advance. So far, \tool has collected more than 5 million Android apps from \numOfSources different sources. These apps and processed data occupy a storage of over 50T, and acquire around 3,472 days in total to process (assuming it takes 1 min to process one app).
  \item \emph{A novel knowledge representation for millions of Android apps.} We make the first attempt to employ a knowledge graph to represent millions of apps and their relationships in between. The apps are computed pairwise and connected based on the similar or identical attributes. The knowledge graph provides an illustrative presentation to show the correlations between apps, and facilitates the extraction of useful facts for research.
  %\item \emph{A comprehensive methodology to extract facts from knowledge graph.} Towards different research problems, we have employed learning algorithms to extract related facts. These facts provides researchers with more concise but precise data that can facilitate the research. In this work by far, \tool has offered \numOfFacts types of facts to our previous studies.
  \item \emph{\tool's practicality in powering up the Android research}. \tool has prepared accurate and abundant data for several research works including malware detection, malware propagation, and app testing. The data of high quality cannot only save researchers' time of data collection, but also facilitate better experiment results.
\end{itemize}

\begin{figure*}
  \centering
  \includegraphics[width=1\textwidth]{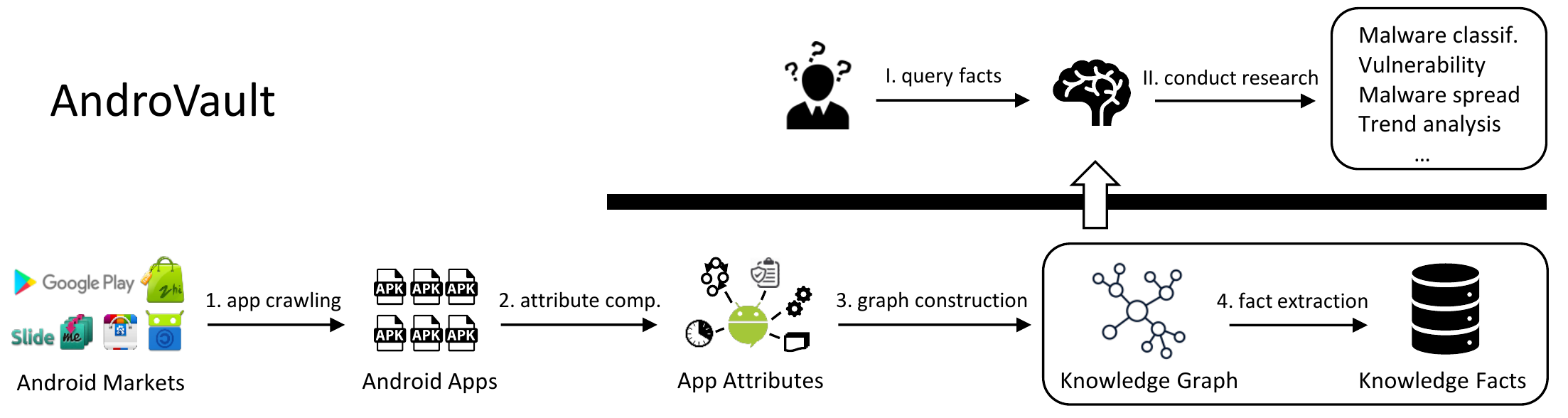}
  \caption{The workflow of AndroVault}\label{fig:arch}
\end{figure*}
\section{The AndroVault Approach}

\textsc{AndroVault} proceeds in four steps as shown in Figure~\ref{fig:arch}. It takes Android app markets including Google Play as input, and produces a knowledge graph combing millions of entities including apps. Moreover, \textsc{AndroVault} employs multiple technologies to produce facts that can facilitate the Android research. For ease of use, \tool provides a computation and clustering engine for researchers which can produce a set of apps of researchers' high interest as well as the corresponding meta information. In particular, we describe each step as follows:
\begin{itemize}
  \item \textbf{App crawling.} \textsc{AndroVault} downloads apps from the official market Google Play and \numOftrdMarkets third-party markets. Besides apk files, we also download their additional information for example app description, number of installation, review score, and category. In addition, we have also collected a large number of apps from publicly known app repositories including \textsc{MalGenome}~\cite{malgenome2012}, \textsc{Drebin}~\cite{drebin2014}, \textsc{VirusShare}\footnote{http://virusshare.com}, \textsc{AMD}~\cite{amd2017} and \textsc{AndroZoo}~\cite{androzoo2016}. %The crawler is implemented on top of \textsc{Scrapy}\footnote{https://scrapy.org/} and \textsc{GoogleplayDownloader}\footnote{https://framagit.org/tuxicoman/googleplaydownloader}.
  \item \textbf{Attribute computation.} Attributes are a condensed and abstract representation, which describes the characteristics of Android apps. There are basically \numOfAttributes types of attributes produced by \textsc{AndroVault} as elaborated in Section~\ref{sec:method:attribute}. Attributes are a bridge to connect or cluster apps with each other. With the computed attributes, researchers can perform machine learning to classify malware, or compute functional similarity. For example, some attributes (e.g., permissions, invoked APIs) can be regarded as feature and passed to a classification algorithm to detect malware.
  \item \textbf{Graph construction.} Android apps lie in the dataset independently, which may cause inconvenience in a large-scale analysis of app. This step is to eliminate the information isolated island of app, and establish connection between apps and relevant data. Inspired by technologies of knowledge representation, we propose two types of relationships between apps, i.e., probabilistic relationship and deterministic relationship. These relationships act on attributes and serve as a property for them (referred to Section~\ref{sec:method:graph} for detail). Eventually, we construct a huge knowledge graph for the entire set of apps, in which nodes are either individual apps with associated attributes or other data (e.g., app author), and edges are connections between these knowledge nodes with respect to a certain relationship. The statistics of the graph is elaborated in Section~\ref{sec:eval:stat}.
  \item \textbf{Fact extraction.} Facts are high-order and composite properties associated with one individual or a set of apps. Generally, facts assist in certain research tasks by providing a prepared data set as well as auxiliary information. They are derived from the built graph, and produced by a variety of technologies such as machine learning, deep learning, logical inference, and statistical computation. For example, representative features for each malware family can be extracted by feature selection technology. These features as signature could be used to classify malware. \tool has computed \numOfFacts types of facts that are beneficial for several research studies. %It is worthy mentioning that they are extendable with new facts that are useful for research studies.
\end{itemize}

So far, we construct a knowledge graph and \numOfFacts types of facts based on it. These knowledge, together with a knowledge query system, provide researchers with ample and precise data as described below.

\noindent\textbf{Application scenarios.} The upper layer in Figure~\ref{fig:arch} shows how researchers utilize the constructed knowledge graph as well as associated facts, and then conduct their research. In particular, researchers query apps or certain facts from \tool, and \tool can produce a list of corresponding apps as well as their attributes by traversing the knowledge graph. Based on these information, researchers can continue with their research such as code smell identification, malware spread, and plagiarism detection. For example, one security analyst wants to identify and then analyze the piggybacked apps, he or she can query the query facts in which apps are of high similarity in code, or have same package name and version code but are signed with different certificates. \tool can produce a set of more representative apps, and thereby facilitate the analyst's research. In another scenario, one deep learning expert intends to study the evolution of code style for one developer, and he or she can query the apps developed by this developer. Based on the selected apps by \tool, this expert is able to quickly employ his/her deep learning algorithm to identify the evolutionary patterns of code, or detect plagiarism with the trained model.

\begin{figure*}
  \centering
  \includegraphics[width=0.9\textwidth]{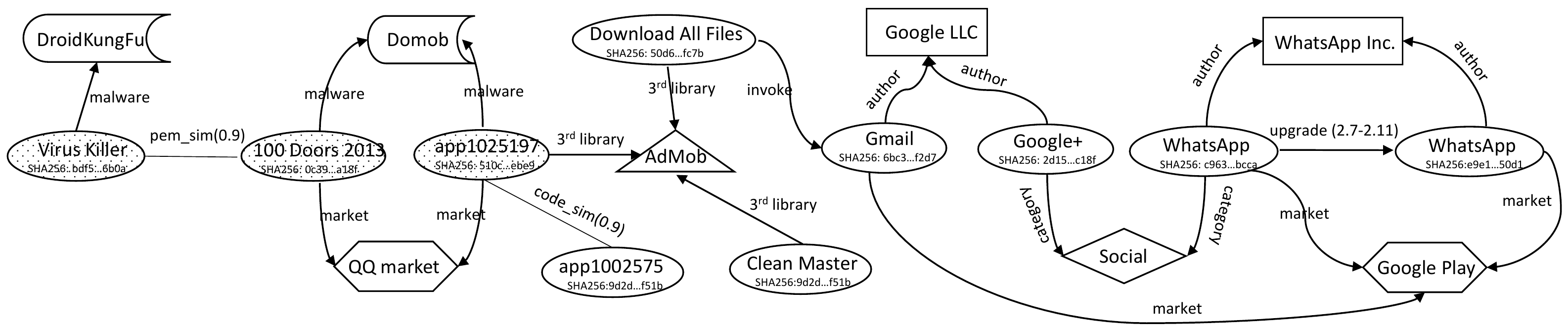}
  \caption{A simplified knowledge graph of \tool}\label{fig:graph}
\end{figure*}

\section{Methodology}\label{sec:method}
In this section, we present the methodologies for each step, and elaborate the output results accordingly.

\subsection{App Crawling}\label{sec:method:crawl}

%\textcolor{red}{Need a workflow graph?}
The input for \emph{app crawling} is a list of Android markets, and \tool contains a web crawler to download apk files periodically. The crawler performs a vertical search limiting its search space within target markets. The crawler is built on top of \textsc{Scrapy} and \textsc{GooglePlayDownloader} for the \numOftrdMarkets third-party markets and Google Play, respectively. %In order to collect up-to-date data, our crawler is designed with three components: \emph{a downloader} to download apk files from markets, \emph{a manager} to start and stop the downloading periodically, and \emph{a delayer} to delay the crawling process to avoid to violate the visit policy of Android markets.

\tool has collected two kinds of information with the crawler: apk files and descriptive information. Apk files are \emph{Android Package Kit}, which is a package format for distribution and installation in Android. \tool has totally collected \numOfApps apk files from \numOfSources different sources. Descriptive information are information that describe the detailed information of apps in our knowledge such as app description, download times, version code, and scores. \tool has totally collected \numOfMetaData types of descriptive information from Google Play.

\subsection{Attribute Computing}\label{sec:method:attribute}

Attributes are representative features that characterize apps. In order to establish the connections between apps, \tool computes \numOfAttributes attributes in total from four aspects as described below.  % to \textsc{AndroZoo++}, \textsc{AndroVault} computes attributes from 5 aspects whic

\noindent\textbf{Manifest file}. We extract the following 11 attributes from the manifest file of app. The attributes in Manifest contain \emph{components} including activities, services, broadcast receivers and content providers, \emph{declared permissions}, \emph{request permissions}, \emph{libraries}, \emph{app name}, \emph{package name}, \emph{version code}, \emph{version name}, \emph{max sdk}, \emph{min sdk}, and \emph{target sdk}.

\noindent\textbf{Dex file}. We perform a lightweight static analysis on app and extract the following 7 attributes: \emph{sha256} that provides an almost-unique, fixed size 256-bit cryptographic hash for app, \emph{certificate} with which apps are digitally signed; \emph{compilation date} denoting the creation date; \emph{invoked Android APIs} in app; \emph{strings}, which are contained in app; ``\emph{centroids}''~\cite{3dcfg2014} for each method in app, and; \emph{invoked apps} with which the app interacts.

\noindent\textbf{Security report}. We fetch \emph{detection results} for each app from \textsc{VirusTotal}, in which there are malware information detected by 57 antivirus tools. In addition, we employ ~\textsc{AVClass} to normalize \emph{family labels} for detected malware. The security report contributes 2 attributes for computing.

\noindent\textbf{Crawl information}. During the crawling process, we also crawl the following \numOfMetaData types of information as attribute: \emph{category}, \emph{description}, \emph{screenshots}, \emph{reviews}, \emph{scores}, \emph{what's news}, \emph{updated date}, \emph{file size}, \emph{install number}, \emph{version}, \emph{required Android version}, \emph{price}, \emph{content rating}, \emph{developer}, \emph{similar apps}, and \emph{market}.

In particular, attributes are of either a primary or composite type. For example, ``package name'' and ``app name'' are of type \emph{string}, while ``request permissions'' and ``detection results'' are of composite type which contain a set of elements. In order to compute attributes from dex file, we implement a lightweight static analyzer to perform a pattern matching in code for attributes ``invoked Android APIs'', and ``strings''. The attribute ``invoked apps'' can be extracted by identifying the outgoing Intents in code.

\subsection{Graph Construction}\label{sec:method:graph}

\tool provides a graph-based knowledge representation for the collected and computed attributes. The knowledge graph is a three tuple, in particular, $G~=~(V, E, \mathcal{R})$.
\begin{itemize}
  \item $V$ is the set of knowledge entities on Android. In addition, $V~\in~\{APP, MARKET, FAMILY, AUTHOR, \\ LIBRARY, CATEGORY\}$, in which one entity could be an app (identified by sha256), an app market (e.g., \textsc{Anzhi}), a malware family (e.g., \textsc{DroidKungFu}), an app author (e.g., WhatsApp Inc.), a third-party library (e.g., \textsc{AdMob}), or an app category (e.g., Social).
  \item $\mathcal{R}$ is the set of relationships which are used to describe the correlation between knowledge entities. There are two types of relationships---\emph{deterministic relationship} ($\mathcal{R}_d$) and \emph{probabilistic relationship} ($\mathcal{R}_s$), i.e., $\mathcal{R}\in\{\mathcal{R}_d, \mathcal{R}_s\}$. In particular, a deterministic relationship defines a certain connection between two entities, while a probabilistic relationship describe a connection between two entities which is probabilistic ranging in (0, 1). For instance, the deterministic relationship ``$author$'' denotes that one knowledge entity is created by another, and $(code\_sim, 0.8)$ means that one knowledge entity is code similar to another with 80\% probability. More relationships are presented in Table~\ref{tbl:relation}.
  \item $E:~V~\times~V~\rightarrow~\mathcal{R}$, which is the set of edges that link two entities by a specific relationship. As shown in Figure~\ref{fig:graph}, the app $Gmail$ is created by the developer $Google LLC$. Hence, there is one link between the entities $Gamil$ and $Google LLC$, i.e., $(Gmail, Google LLC)~\rightarrow~(author)$. The malware $app1025197$ has similar code with the malware $app1002575$, so there is a probabilistic relationship in between as $(app1025197, app1002575)~\rightarrow~(code\_sim, 0.9)$.

\end{itemize}

Figure~\ref{fig:graph} shows a simplified knowledge graph combining all types of knowledge entities and a partial set of relationships. In detail, we can easily draw some conclusions from this graph. For example, the apps Google+ and WhatsApp belong to the same category; the two apps with different hash values (c963...bcca, e9e1...50d1) are all WhatsApp but of different versions; two apps app1025197 and app1002571 created by one author are similar (0.9) in code to each other; and the app ``Download All Files'' has invoked the app ``Gmail'' by sending it an Intent.

To construct a knowledge graph for Android is to compute all relationships listed in Table~\ref{tbl:relation}. The deterministic relationships are established with the identical attributes of apps. %Specifically, the ``invoke'' relationship is computed by a lightweight static analyzer to identify a behavior of sending an Intent to another app.  
The probabilistic relationships are computed by employing different similarity metrics between apps or markets as follows.

\begin{figure}
  \centering
  \includegraphics[width=0.45\textwidth]{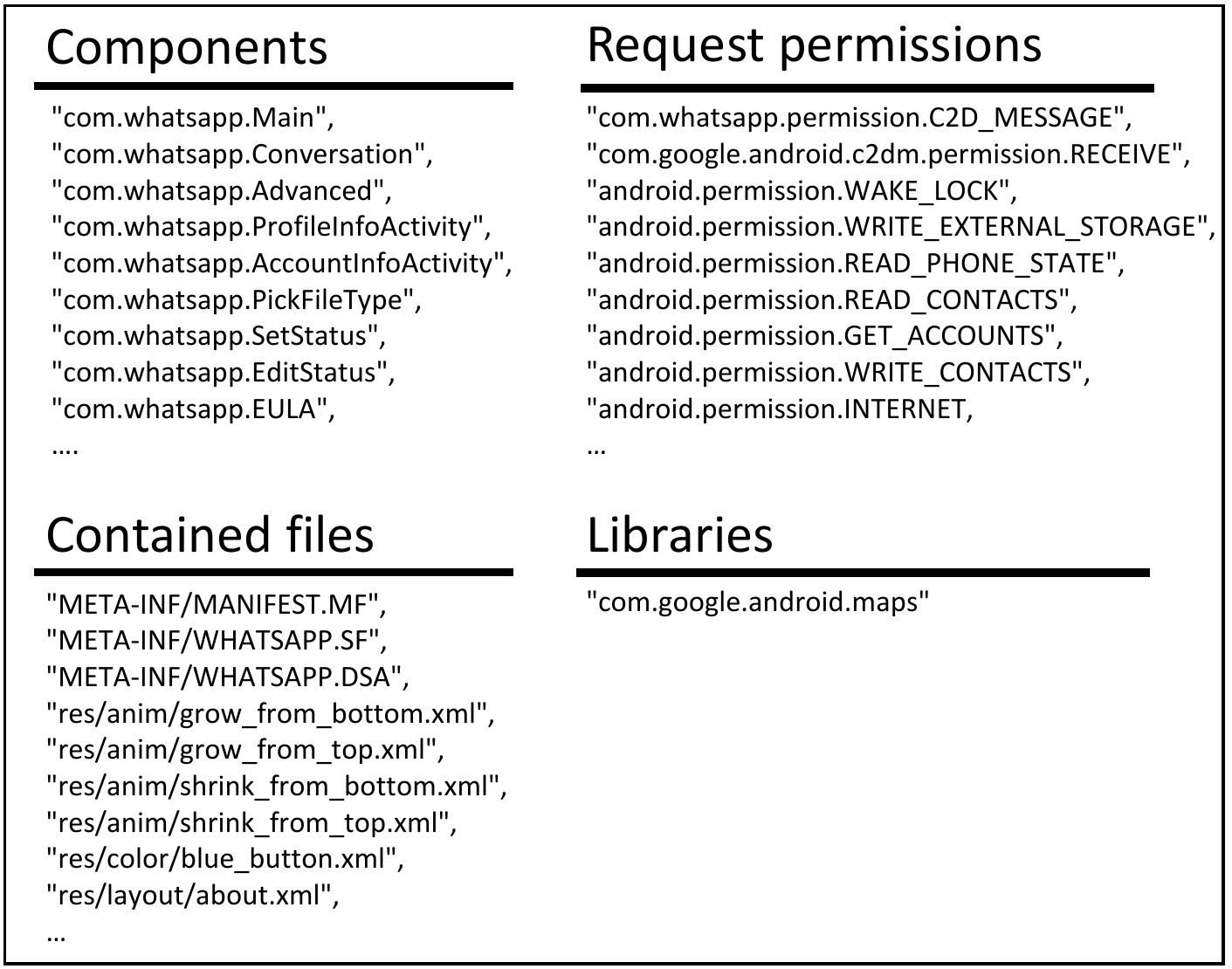}
  \caption{Composite attributes on which Jacaard is employed}\label{fig:composite_attr}
\end{figure}

\noindent\textbf{Jacaard Index}. We employ \emph{Jacaard Index} to compute the similarity of the composite attributes that exhibit a set of elements without any particular order. As shown in Figure~\ref{fig:composite_attr}, four types of composite attributes of apps contain a set of strings, and are compared with the Jacaard Index.

\noindent\textbf{3D-CFG Similarity}. To compute code similarity between apps, we employ the approach in~\cite{3dcfg2014,smart2016} to first compute the centroid for each Java method, compute the distance between methods pairwise, and eventually obtain the distance between apps.

To better depict the proximity between knowledge entities and reduce the size of the knowledge graph, we only retain the probabilistic relationships with a probability above 0.9. More results are presented in Section~\ref{sec:eval:stat}.

\subsection{Fact Extraction}\label{sec:method:fact}

Facts are high-order and composite properties associated with one individual or a set of apps. Generally, facts support a certain research study by providing either a prepared data set and associated attributes. Many technologies can be applied to extract facts such as machine learning, statistical inference, and graph-based searching. %For example, \tool assists a study of piggybacked apps by producing a number of apps that are modified by another author to original apps. 
In particular, we present \numOfFacts types of facts produced by \tool produces as follows to support different research studies. It is worthy mentioning that \tool can be extended to extract more types of facts to aid in other research problems.

\noindent\textbf{Malicious Code Localization}. Generally, malware samples in the same family share common malicious behavior. In such a case, \tool performs a clustering algorithm to identify methods that are representative for this family. These methods are where malicious code is located, and could be represented as signature for malware detection.  The fact has powered up two projects~\cite{smart2016,onlinelearning2017}.

\noindent\textbf{Market Correlation}. Apps are commonly copied between markets. It is necessary to learn the copy mechanism for the study of malware spread between markets. \tool computes the replication ratio for each market, and identify the commonality between markets. The result has powered up one study of malware spread between app markets.

\noindent\textbf{Suspicious App Identification}. An update attack is conducted by one author adding malicious code in his/her old-fashioned apps. \tool identifies these suspicious apps by searching the knowledge graph with three attributes (i.e., package name, detection result, and certificate) of app. A piggybacked app is created by another author modifying codes in the original app. \tool identifies piggybacked apps by searching the knowledge graph with three attributes (i.e., package name, version code, and certificate).

\begin{figure}
  \centering
  \includegraphics[width=0.35\textwidth]{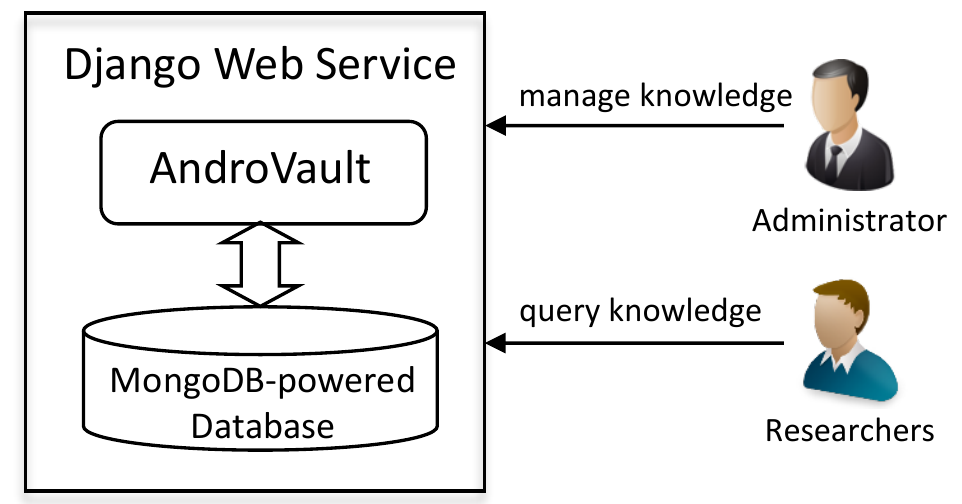}
  \caption{System architecture for the \tool service}\label{fig:system}
\end{figure}
\section{Tool Implementation \& Results}

\textsc{AndroVault} is implemented with \numOfJava lines of Java and \numOfPython lines of Python. In addition, it employs several state-of-the-art tools to achieve specific tasks: \textsc{Scrapy}\footnote{https://scrapy.org/} and \textsc{GoogleplayDownloader}\footnote{https://framagit.org/tuxicoman/googleplaydownloader} contribute the implementation of our crawler to download Android apps from Google Play and third-party markets; \textsc{AndroGuard} and \textsc{Soot}~\cite{soot1999} are used to perform a static analysis on app; \emph{AVClass} is used to standardize the name of malware families; and \emph{scikit-learn}\footnote{http://scikit-learn.org/stable/} is used to perform machine learning tasks to extract new knowledge.

\begin{figure}
  \centering
  \includegraphics[width=0.5\textwidth]{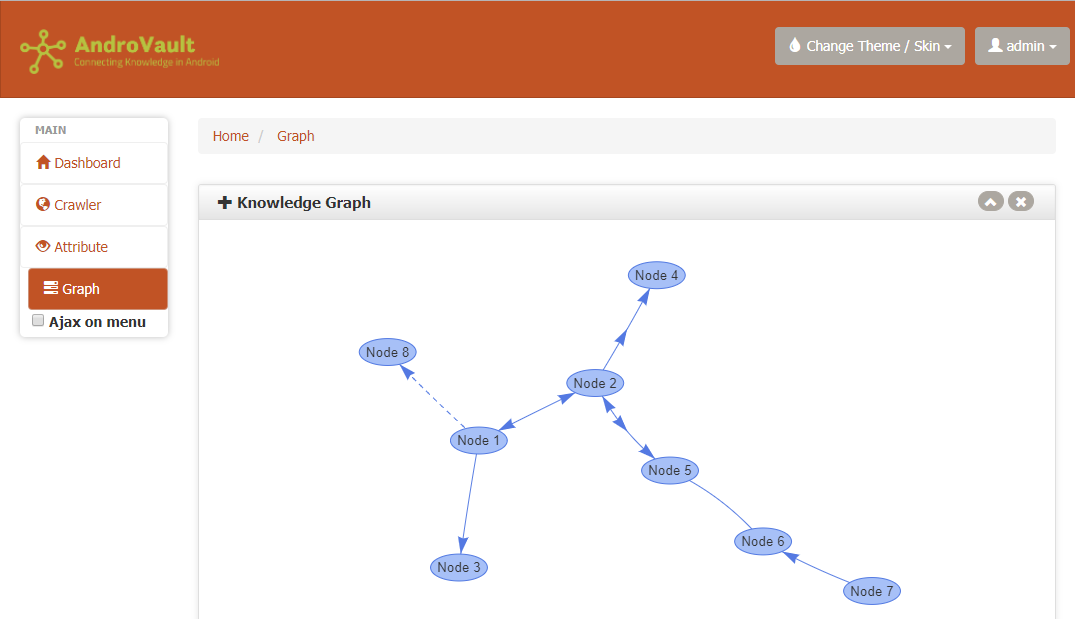}
  \caption{A screenshot for the visualization system}\label{fig:screenshot}
\end{figure}

For ease of use, we build a web service to manage \tool and visualize the knowledge as shown in Figure~\ref{fig:system}. The web service is based on \textsc{Django}\footnote{https://www.djangoproject.com/} which responds the requests from the administrator and researchers in need.  The knowledge produced by \tool is stored in \textsc{MongoDB}, a NoSQL database. Benefitting from the handy and various functions provided by this web service, the administrator can manage the \tool engine with its substantial functions of data collection, knowledge representation and fact extraction. Android researchers are able to query knowledge, and obtain their requisite data and knowledge to accomplish their goals. Figure~\ref{fig:screenshot}, a screenshot of the web service, illustrates the GUI to visualize the knowledge graph.

\subsection{Statistics of Knowledge Graph}\label{sec:eval:stat}

In this section, we present the statistics of collected and produced data by \tool.

\noindent\textbf{Collected Data.} Since 2013, we have crawled Android apps and other unstructured data from the Internet for four years. \tool integrates \numOfMarkets crawlers to download apps from corresponding markets, and one additional crawler to download app description from Google Play. To date, we have collected \numOfApps Android apps, and \numOfAppDesc entries of app description.% (in which XX\% are for Android apps, and the remaining ones are for Apple apps).

\begin{figure}
  \centering
  \includegraphics[width=0.5\textwidth]{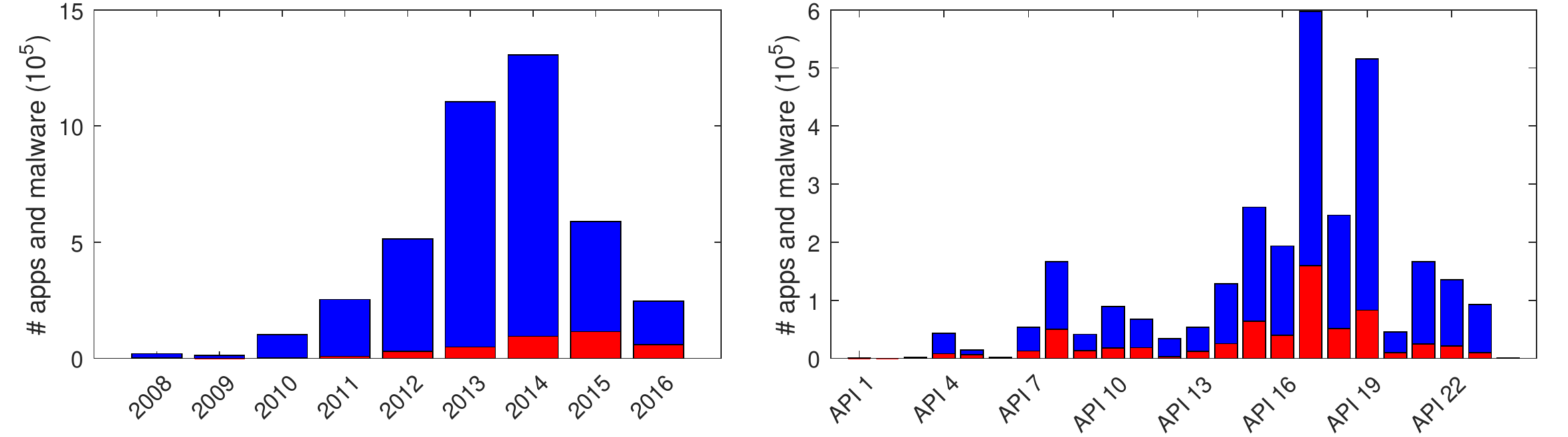}
  \caption{Statistics of collected apps and malware. The x-axis of the left figure is the years since the malware is created, and the y-axis denotes the number of apps and malware. The x-axis of the right figure is the Android versions from API 1 to API 24, and the y-axis denotes the number of apps and malware. The blue bar denotes the number of apps, and the red bar denotes the number of malware.}\label{fig:statistics}
\end{figure}

We present all markets and the statistics of collected data in Table~\ref{tbl:market_description}. In addition, we download \numOfOtherApps apps from four well-known app repositories: \textsc{MalGenome}, \textsc{Drebin}, \textsc{VirusShare}, and \textsc{AndroZoo}. According to \textsc{VirusTotal}'s detection results, we determine one app to being malware by the confirmation of at least one antivirus tool. In Figure~\ref{fig:statistics}, we plot the distributions of apps and malware over time and Android versions.

\noindent\textbf{Graph Presentation.} We present the statistics for the entities \emph{market}, \emph{family}, \emph{author}, and \emph{category} in the knowledge graph. Table~\ref{tbl:market_description} displays the description for the 28 app markets including location, language, hosted country, number of apps, and number of malware.

\begin{table}
\begin{center}
\caption{Top 10 families with the most number of malware}\label{tbl:family}
\resizebox{0.45\textwidth}{!}{
\begin{tabular}{lrl}\toprule
\textbf{Family} & \textbf{Number} & \textbf{Malicious Behaviors} \\ \midrule
kuguo & 25,002 & privacy (e.g., location), display ads, install apps\\
airpush & 17,135 & privacy (e.g., IMEI), create shortcut or bookmarks\\
dowgin & 16,818 & privacy (e.g., IMEI), display ads, exec. payloads\\
smsreg & 10,737 & privacy (e.g., location, IMEI, network operator)  \\
secapk & 9,885 & code is protected by an app shield\\
gappusin & 9,452 & download payloads, display ads, install shortcut\\
revmob & 8,731 & privacy (e.g., account, IMEI), display ads\\
leadbolt & 6,636 & change settings, display ads\\
youmi & 5,919 & privacy (e.g., IMEI), display ads, install apps \\
domob & 5,718 & privacy (e.g., IMEI), display ads, install apps\\ \bottomrule
\end{tabular}
}
\end{center}
\end{table}

\textsc{AndroVault} obtains 1,004,550 malware samples by virtue of \textsc{VirusTotal}, and further classifies 200,373 of them into 1,400 families using \textsc{AVClass}. As shown in Table~\ref{tbl:family}, we present ten families with the most number of malware, as well as the contained malicious behaviors.

Apps have to be digitally signed before release with a specific certificate. Therefore, the certificate of apps can be used to recognize different authors. \tool extracts the information of certificate for all apps, including the public key, the detailed information of issuer, and the detailed information of subject. We present the distribution of authors of identified malware in Table~\ref{tbl:authorship}. The first column refers to the authors which produce malware in a range of number. For example, ``1-10'' denote the number of the authors' malware we can find in the dataset is in the range from 1 to 10; the second column refers to the percentage of the authors over all malware authors, and; the third column refers to the percentage of the malware written by these authors over all malware samples.

\begin{table}[t]
\begin{center}
\caption{The distribution of authorship of Android malware. }\label{tbl:authorship}
\begin{tabular}{lrrrr}\toprule
\multirow{2}{*}{\# Range} & \multicolumn{2}{c}{\textbf{Author}} & \multicolumn{2}{c}{\textbf{Total Malware}} \\ \cline{2-5}
& \# & \% & \# & \% \\ \hline
%\textbf{\# Malware}& \textbf{\% Author} & \textbf{\% Malware} \\ \hline
1-10    & 223,702 & (96.2\%) & 374,296 & (39.1\%)\\ %\hline
11-100  & 8,116 & (3.5\%) & 211,169 & (22.0\%)\\% \hline
101-500 & 600 & (0.3\%) & 114,141 & (11.9\%) \\ %\hline
501-1000 & 52 & (--\%) & 35,208 & (3.7\%) \\ %\hline
$>$1000   & 65 & (--\%) & 222,874 & (23.3\%) \\ \bottomrule
\end{tabular}
\end{center}
\end{table}

\begin{figure}
  \centering
  \includegraphics[width=0.45\textwidth]{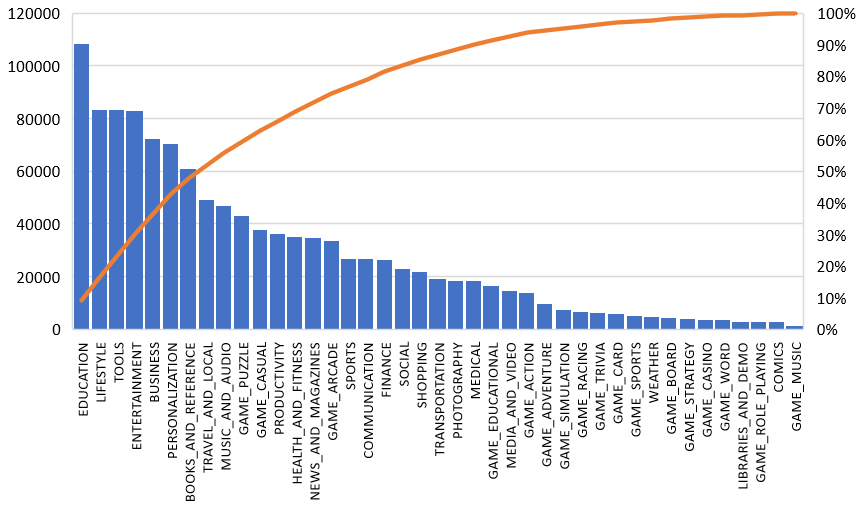}
  \caption{App distribution in categories}\label{fig:category}
\end{figure}

Apps are labeled with a unique category in Google Play for a brief description. \tool incorporates 41 categories, and classifies \numOfOtherApps apps into these categories. Figure~\ref{fig:category} shows the distribution of apps in terms of category.%As shown in Figure~\ref{fig:category}, the categories \emph{education}, \emph{lifesyle}, \emph{tools}, \emph{entertainment} and \emph{business} occupy the first five places.

\noindent\textbf{Extracted Facts}. \tool has extracted a number of facts that could power some research studies. In a nutshell, \tool computes all shared apps between markets, and finds that market \textsc{F-Droid} has the maximal portion (92.4\%) of replicated apps, and market \textsc{AppFun} has the minimal portion (1.0\%) of replicated apps. The detailed data of sharing between markets are beneficial for the study of app spread between markets.

\tool identifies \numOfPackages apps by their package names (\numOfAppInPackages apps in total). Each app has \numOfVersions versions on average. Moreover, we find that \numOfCorruptedApps (14.4\%) of them become malware during the upgrade process, and \numOfPiggybackedApps (29.3\%) of them are modified by another author which are likely piggybacked apps. The facts could be further utilized for app corruption analysis (why apps become malicious during upgrade), and feature extraction for piggybacked apps.

\tool employs 3D-CFG similarity to identify cloned code for 149 malware families in \textsc{Drebin}. The cloned code can be used as signature to give a find-grained representation for each family. In total, \numOfMethods methods are extracted, and each family has 49 methods that represent malicious behaviors on average.

\subsection{Use Cases}

The collected data and extracted knowledge have driven a variety of substantial studies. In this section, we elaborate three of them to demonstrate the efficiency and practicability of \tool.

\noindent\textbf{Malware Detection.} Machine learning is commonly used to detect malware, and an accurate, representative, and updated data set of Android apps and malware is very significant to detection performance. \tool provides 44,347 benign apps and 42,910 malware in a span of 224 days for~\cite{onlinelearning2017}. In addition to the selected malware over time, \tool also provides several attributes of them such as requested permissions. Based on the data, the study is superior in detecting evolving malware that exploits new vulnerabilities or attack technologies.

\noindent\textbf{Malware Propagation.} Malware propagation between markets on Android has not been thoroughly studied before. One of the reasons is that the detailed data depicting the propagation of malware within or between markets is absent for researchers. \tool supports this study (i.e., malware spread between markets) by providing the distributions of malware within markets, and detailed family labels of these malware samples. Based on the prepared data, this study can identify the growth model within markets, and the spread model between markets. Further, it can facilitate the studies of assessing the security of markets, and identifying the infection capability of malware.

\noindent\textbf{Android App Testing.} \tool can benefit Android app testing from two aspects: providing entry points for apps that could trigger the behaviors inside the apps. For example, one app may execute specific actions once received a broadcast message. In such a case, this app has to register a broadcast receiver to listen to this message, either in the manifest file, or in code. \tool elaborates the apps to test with these entry points which can be employed by a tester to increase the test coverage; selecting a representative data set for testing in terms of multiple dimensions including popularity, category, and compatibility.

\section{Discussion}\label{sec:discuz}

In this section, we present the applicable areas of \tool, and the future work.
\subsection{Application}

Since \tool is able to provide a more accurate data set according to researchers' requirements, there are many potential applicable areas where \tool can benefit. In the following, we detail its four applicable areas with answering how \tool aids in them.

\noindent\textbf{Feature Evaluation}. Feature selection is an important yet challenging task in machine learning. The quality of features determines the performance of machine learning approaches to a large extent. Good features are supposed to be concise, representative, and non-overfitting, which require extensive experiments to evaluate. This problem also perplexes Android researchers since they are using a large number of machine learning techniques to solve issues on Android, such as malware detection, third party library detection, family labelling, and app recommendation. Given by \tool \numOfAttributes types of features for data on Android, it allows researchers to conduct an immediate experiment to evaluate the features for different research goals. In addition, feature selection algorithms can also be performed on these well prepared features, and identify most suitable features for a specific research task.

\noindent\textbf{Trend Analysis}. It has become a hot topic of trend analysis in the Android field, for example, how malware evolves during the endless war against security defenders, and how code is refactored to improve its security, performance, modularity, and readability during its sustainable upgrade process. \tool records and maintains abundant attributes for apps to facilitate this kind of research. Therefore, one clustering algorithm can be immediately used to extract a set of data of the most interest. If, for example, researchers are working on a study of malware evolution, they can utilize \tool to produce a list of malware samples that belong to a same family while share different characteristics (e.g., requested permission, invoked APIs and dependent libraries) according to specific requirements. This can spare researchers' time to collect and purify related data, which may not the main contributions in their research.

\noindent\textbf{Risk Assessment}. It is well known that we can assess the risk of apps by checking whether it contains unwanted or malicious behaviors. Fortunately, researchers can accomplish a more comprehensive assessment with the help of \tool. As we know, it has been commonly applied in assessing the risk or trustworthiness of one node by the opinions of its neighbors in reputation system. In a similar way, we can build a reputation system on the constructed knowledge graph by \tool. In the constructed knowledge graph, researchers can choose suitable rating data  from all present relationships between apps. Based on the rating data, they manage to compute a quantitative value for  risk. The relationships, such as the shared third party libraries, similar reviews given by users, and the common developers , are all potentially useful to assess the risk of apps.

\noindent\textbf{Collusive Attacks}. Collusive attacks mean one kind of attacks that are completed by two or more involved apps, which are widely occurring on Android~\cite{dialdroid2017}. However, it is a time-consuming task to identify the collaborative apps in an enormous number of apps pairwise. Since the Intent via Binder is the most prominent method for inter-application communication, \textsc{AndroVault} employs a lightweight static analysis to identify possible communication behaviors existing in code, and builds a connection between these apps in communication. Therefore, researchers can leverage this data to perform a more accurate analysis of the communication, and further confirm whether one collusive attack occurs.

\subsection{Future work}

There are still some promising research directions of this work. We are going to incorporate code feature with natural language description, and build a correlation in between. For example, one app may claim in its description that it does not read a file in the external storage, however, it breaks the vow and stealthily access sensitive files. This provides a strong clue that the app may be conducting malicious behaviors stealthily. Therefore, the fidelity information (i.e., what it behaves is what is described (WIBIWID)) can benefit several researches such as malware detection, consistency checking, and code profiling.

Execution trace of apps can exactly reveal their behaviors during execution. Some apps may behave differently on real devices and virtual machines in order to avoid the security inspection by analysts. Other apps may exploit exposed vulnerabilities to escape the Android Sandbox and gain privileged permissions. In such scenarios, it is desirable to collect dynamic execution traces of apps and perform an in-depth analysis on them. Due to the time cost of dynamic testing (e.g., one app costs three hours for testing), we are unable to collect many dynamic execution traces currently. We intend to continue this collection work and produce more knowledge on the data.

\tool are also dedicated to tackling vulnerability issues on Android. A large proportion of exploits are implemented in native code, and moreover, there are a massive volume of web-based resources profiling these vulnerabilities like National Vulnerability Database (https://nvd.nist.gov/). In the future, we intend to collect these web-based resources, and compute attributes of native code in apps. The resulting knowledge can facilitate the detection of vulnerabilities and exploits. 
\section{Related Work}\label{sec:related}

%\emph{Knowledge　Vault} \cite{knowledgevault2014} is a web-scale approach to construct a probabilistic knowledge on the basis of \emph{Knowledge Graph}~\cite{knowledgegraph2013}. It extracts information triples from Web, and employs supervised machine learning methods to fuse with structured facts in Knowledge Graph. The probabilities in the base denote the likelihood of fact correctness. 

\textsc{MalGenome}~\cite{malgenome2012} is an initiative study to dissect Android malware and its evolution. It provides a considerate number (1,260) of malware samples to the public with detailed malicious behaviors inside. Subsequently, \textsc{Drebin}~\cite{drebin2014} aggregate 5,560 malware samples which are identified and classified by a machine learning method. These samples are categorized into 149 families in terms of contained malicious behaviors. 

Wei~\textit{et al.} leverage the detection results from \textsc{VirusTotal} and similarity computation to produce find-grained labels for Android malware~\cite{amd2017}. Furthermore, they conduct an in-depth manual analysis to present the detailed malicious behaviors inside the malware. The final dataset~\textsc{AMD} currently contains 24,553 samples, categorized in 135 varieties among 71 malware families. 

\textsc{AndroZoo}~\cite{androzoo2016} and its extension \textsc{AndroZoo++}~\cite{androzoo2017} have collected over five million Android apps from diverse sources including Google Play to date. In addition, they figure out 20 types of app metadata to share with the research community for relevant research works.

Further to their work, we build a huge knowledge graph based on these attributes by machine learning or statistical inferring correlation between apps. Therefore, we are able to extract some primary knowledge that are extracting from the knowledge graph. In addition, other graph-based approaches are also be employed to produce knowledge in research. 
\section{Conclusion}
In this work, we propose \tool to collect, represent, and abstract knowledge for automated computing. \tool can produce a demanding and accurate data for a specific research task by computing the characteristics of Android apps as well as correlations in between. Based on the data produced by \tool, we have conducted a variety of research works including malware detection, code generation, and Android testing.

\tool is a continuous work, and will be further extended and enhanced with more convenient functions for researchers. For example, we are going to compute more attributes of Android apps related to vulnerabilities, to facilitate the detection of vulnerabilities, and automated exploit generation.

\bibliographystyle{abbrv}
\bibliography{andromine}
%\newpage
%\section*{Appendix}\label{sec:appendix}

\begin{table*}[!h]
\begin{center}
\caption{All relationships between knowledge nodes}\label{tbl:relation}
\begin{tabular}{lll} \toprule
\textbf{Type} & \textbf{Relationship} & \textbf{Description} \\ \midrule
\multirow{7}{*}{\shortstack{Deterministic \\ Relationship}} & malware & the relationship between one app and one malware family denotes the app belongs to this family. \\ \cline{2-3}
&author & the relationship between one app and one author denotes the app is created by this author. \\ \cline{2-3}
&library & the relationship between one app and one library denotes the app contains this library. \\ \cline{2-3}
&market & this relationship between one app and one app market denotes the app is from this market. \\ \cline{2-3}
&category & this relationship between one app and category denotes the app is classified into this category. \\ \cline{2-3}
&upgrade & this relationship between two apps denotes they are of same app but different version. \\ \cline{2-3}
&invoke & this relationship between two apps denoting that one app invokes the another. \\ \midrule

\multirow{7}{*}{\shortstack{Probabilistic \\ Relationship}}&code\_sim & this relationship between two apps denotes their similarity in code. \\ \cline{2-3}
&api\_sim & this relationship between two apps denotes their similarity in invoked APIs. \\ \cline{2-3}
&perm\_sim & this relationship between two apps denotes their similarity in permission. \\ \cline{2-3}
&comp\_sim & this relationship between two apps denotes their similarity in components. \\ \cline{2-3}
%&perm\_sim & this relationship between two apps denotes their similarity in permission. \\ \cline{2-3}
&lib\_sim & this relationship between two apps denotes their similarity in requested hardware features. \\ \cline{2-3}
&file\_sim & this relationship between two apps denotes their similarity in contained files. \\ \cline{2-3}
&mark\_sim & this relationship between two markets denotes their commonality in apps. \\ \bottomrule
\end{tabular}
\end{center}
\end{table*}

\begin{table*}[!ht]
\begin{center}
\caption{The description of the 28 app markets of our crawled apps}\label{tbl:market_description}
\begin{tabular}{llllrr}
\toprule
\textbf{Name}   & \textbf{URL} & \textbf{Language} & \textbf{Country} & \textbf{\# apps} & \textbf{\# malware}  \\ \midrule
%androzoo & \url{https://androzoo.uni.lu/} & English & Global &  2,727,911 & 612,584 \\ \hline
googleplay & \url{https://play.google.com/store?hl=en} & English & Global & 1,252,681 & 84,779 \\ \hline
qq & \url{http://sj.qq.com/myapp/} & Chinese & China & 303,372 & 85,546 \\ \hline
anzhi & \url{http://anzhi.com/} & Chinese & China & 256,495 & 59,570 \\ \hline
getjar & \url{http://www.getjar.com/} & English & Europe & 140,528& 22,535 \\ \hline
mumayi & \url{http://www.mumayi.com/} & Chinese & China & 45,399 & 21,219 \\ \hline
xiaomi & \url{http://app.mi.com/} & Chinese & China & 36,453 & 20,868 \\ \hline
apk20 & \url{http://www.apk20.com/} & English & US & 33,555 & 5,260 \\ \hline
hiapk & \url{http://www.hiapk.com/} & Chinese & China & 29,291 & 11,555 \\ \hline
eoemarket & \url{http://www.eoemarket.com/} & Chinese & China & 27,012 & 12,257 \\ \hline
%virusshare & \url{https://virusshare.com/} & Global & Global & 24,322 & 24,322 \\ \hline
appchina & \url{http://www.appchina.com/} & Chinese & China & 14,769 & 8,858 \\ \hline
baidu & \url{http://shouji.baidu.com/}  & Chinese & China & 13,909 & 425 \\ \hline
coolapk & \url{http://coolapk.com} & Chinese & China & 13,851 & 2,754 \\ \hline
apkmirror & \url{http://www.apkmirror.com/} & English & US & 11,641 & 404 \\ \hline
flyme & \url{http://app.flyme.cn/} & Chinese & China & 11,278 & 4,512 \\ \hline
gfan & \url{http://apk.gfan.com/} & Chinese & China & 10,746 & 5,533 \\ \hline
cnmo & \url{http://app.cnmo.com/} & Chinese & China & 10,745 & 3,831 \\ \hline
androiddrawer & \url{http://www.androiddrawer.com/} & English & Global & 9,106 & 747 \\ \hline
wangyi & \url{http://m.163.com/android/index.html} & Chinese & China & 7,054 & 3,447 \\ \hline
anruan & \url{http://www.anruan.com/} & Chinese & China & 6,781 & 3,233 \\ \hline
%drebin & \url{https://www.sec.cs.tu-bs.de/~danarp/drebin/}  & Global & Global & 5,560 & 5,560 \\ \hline
slideme & \url{http://slideme.org} & English & US & 4,711 & 209 \\ \hline
fdroid & \url{https://f-droid.org/} & English & US & 3,757 & 64 \\ \hline
freewarelovers & \url{http://www.freewarelovers.com/android} & English/German & Germany & 3,595 & 248 \\ \hline
mob & \url{http://mob.org/} & English & US & 2,914 & 567 \\ \hline
wandoujia & \url{http://www.wandoujia.com/apps}  & Chinese & China & 2,888 & 1,641 \\ \hline
apkpure & \url{https://apkpure.com} & English & US & 2,827 & 435 \\ \hline
appsapk & \url{http://appsapk.com} & English & US & 1,997 & 196 \\ \hline
huawei & \url{http://appstore.huawei.com/} & Chinese & China & 1,771 & 0 \\ \hline
%genome & \url{http://www.malgenomeproject.org/} & Global & Global & 1,260 & 1,260 \\ \hline
chinamobile & \url{http://mm.10086.cn/} & Chinese & China & 236 & 131 \\ \hline\hline
%angeeks & \url{http://www.angeeks.com/} & Chinese & China & ? & ? \\ \hline
%appfun & \url{http://appfun.cn/} & Chinese & China & ? & ? \\ \hline
\textbf{Total} & & & & 2,246,616 & 360,824 \\ \bottomrule
\end{tabular}
\end{center}
\end{table*} 
\end{document}